\documentclass[reprint, aps, prx, floatfix, superscriptaddress]{revtex4-2}
\usepackage{graphicx}    
\usepackage{dcolumn}    
\usepackage{bm}             
\usepackage{amssymb}   
\usepackage{amsmath}    
\usepackage{amsthm}
\usepackage{commath}   
\usepackage{subfigure}    
\usepackage{braket}
\usepackage{natbib}
\usepackage{float}
\usepackage{color}
\usepackage{siunitx}
\usepackage[colorlinks=true, allcolors=blue]{hyperref}
\usepackage{hyphenat}
\usepackage{notes2bib}
\usepackage{MnSymbol}

\graphicspath{{Figures/}}

\newtheorem{theorem}{Theorem}

\begin{document}

\title{Information transmission with continuous variable quantum erasure channels}

\author{Changchun Zhong}
\email{zhong.changchun@uchicago.edu}
\affiliation{Pritzker School of Molecular Engineering, University of Chicago, Chicago, IL 60637, USA}

\author{Changhun Oh}
\affiliation{Pritzker School of Molecular Engineering, University of Chicago, Chicago, IL 60637, USA}

\author{Liang Jiang}
\affiliation{Pritzker School of Molecular Engineering, University of Chicago, Chicago, IL 60637, USA}

\date{\today}

\begin{abstract}

Quantum capacity, as the key figure of merit for a given quantum channel, upper bounds the channel's ability in transmitting quantum information. Identifying different types of channels, evaluating the corresponding quantum capacity, and finding the capacity-approaching coding scheme are the major tasks in quantum communication theory. Quantum channel in discrete variables has been discussed enormously based on various error models, while error model in the continuous variable channel has been less studied due to the infinite dimensional problem. In this paper, we investigate a general continuous variable quantum erasure channel. By defining an effective subspace of the continuous variable system, we find a continuous variable random coding model. We then derive the quantum capacity of the continuous variable erasure channel in the framework of decoupling theory. The discussion in this paper fills the gap of a quantum erasure channel in continuous variable setting and sheds light on the understanding of other types of continuous variable quantum channels.

\end{abstract}

\maketitle

\section{Introduction}
Besides storing and manipulating quantum information, efficiently transmitting them through space and time with high fidelity is essential for many practical quantum applications \cite{hayashi2014, watrous2018,gyongyosi2018,bennett1998}. In quantum communication theory, information transmission is modeled as a quantum channel, which is defined by a completely positive and trace preserving (CPTP) map $\mathcal{N}:\rho_A\rightarrow\rho_B$, where the map $\mathcal{N}$ takes some input $\rho_A\in\mathcal{H}_A$ to the output $\rho_B\in\mathcal{H}_B$ \cite{busch2016}. In general, practical quantum channels are noisy, which means that the output state cannot carry the same amount of information as the input \cite{holevo1998,barnum1998,lloyd1997}. Given a quantum channel, an important task is to find its \textit{quantum capacity}, which gives the ultimate limit of the achievable information transmission rate using the channel. 

In practice, the exact expression of quantum capacity is generally hard to find \cite{gyongyosi2018,eisert2005,devetak2005,holevo2019}, thus a great deal of effort is devoted to improving its upper or lower bound \cite{rosati2018,sharma2018,jeong2020}. Quantum erasure channel is one of the few quantum channels whose quantum capacity is known. By definition, the quantum erasure channel $\mathcal{N}_p$ erases the input state $\rho$ with probability $p$ while keeping the input state unchanged with chance $1-p$ \cite{grassl1997,bennett1997},
\begin{equation}
    \mathcal{N}_p(\rho)\rightarrow (1-p)\rho+p\ket{e}\bra{e},
\end{equation}
where $\ket{e}$ is a flag state that is orthogonal to the input Hilbert space. Denote $d$ as the dimension of the input space, the erasure channel has a quantum capacity $Q=\mathrm{max}\{(1-2p)\log_2d,0\}$ \cite{bennett1997}, which means positive information transmission rate is achievable if the erasure probability is less than half. 

As we noticed, since the first appearance of quantum erasure channel \cite{grassl1997}, almost all discussions are based on discrete variables (DV), while less is investigated in the continuous variable (CV) setting. As CV quantum information processing shows more and more potentials \cite{braunstein2005,weedbrook2012,gottesman2001,ma2021}, it is important to identify the CV channels or the corresponding error models. In Ref.~\cite{niset2008}, a CV model of ``quantum erasure channel" is first considered, which either transmits a perfect coherent state or replaces the input by a vacuum, denoted as $\mathcal{N}(\ket{\alpha}{\bra{\alpha}})=(1-p)\ket{\alpha}\bra{\alpha}+p\ket{0}\bra{0}$. 

\begin{figure}[t]
\centering
\includegraphics[width=\columnwidth]{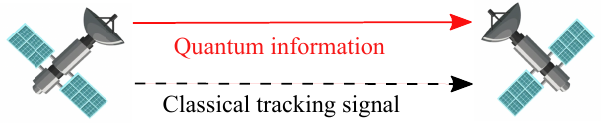}
\caption{The quantum erasure channel realized by sending classical tracking signal along with the quantum information. The detection of the classical signal heralds the quantum information transmission. \label{fig0}}
\end{figure}

Although interesting, this CV model does not faithfully capture the essence of an erasure channel {since the coherent state and vacuum state are not orthogonal.} In this paper, we study the CV quantum erasure channel more rigorously, {where an arbitrary CV input state is either completely erased with a chance $p$ or perfectly transmitted with a probability $1-p$, given as
\begin{equation}
    \mathcal{N}(\rho_\mathrm{cv}\otimes\ket{\uparrow}\bra{\uparrow})\rightarrow(1-p)\rho_\mathrm{cv}\otimes\ket{\uparrow}\bra{\uparrow}+p\ket{0}\bra{0}\otimes\ket{\downarrow}\bra{\downarrow},
\end{equation}
where the spin state is used as a flag to signal the erasure and is important to rigorously define the quantum erasure channel.} {Moreover, this channel is also practical which can be realized by introducing classical tracking signals along the quantum encoding. The detection of classical signal at the receiving end is used to herald the quantum information transmission, as sketched in Fig.~\ref{fig0}. Note that the technique with classical tracking signals is already used in heralding connections between quantum satellites \cite{sidhu2021}.}

\begin{figure}[t]
\centering
\includegraphics[width=\columnwidth]{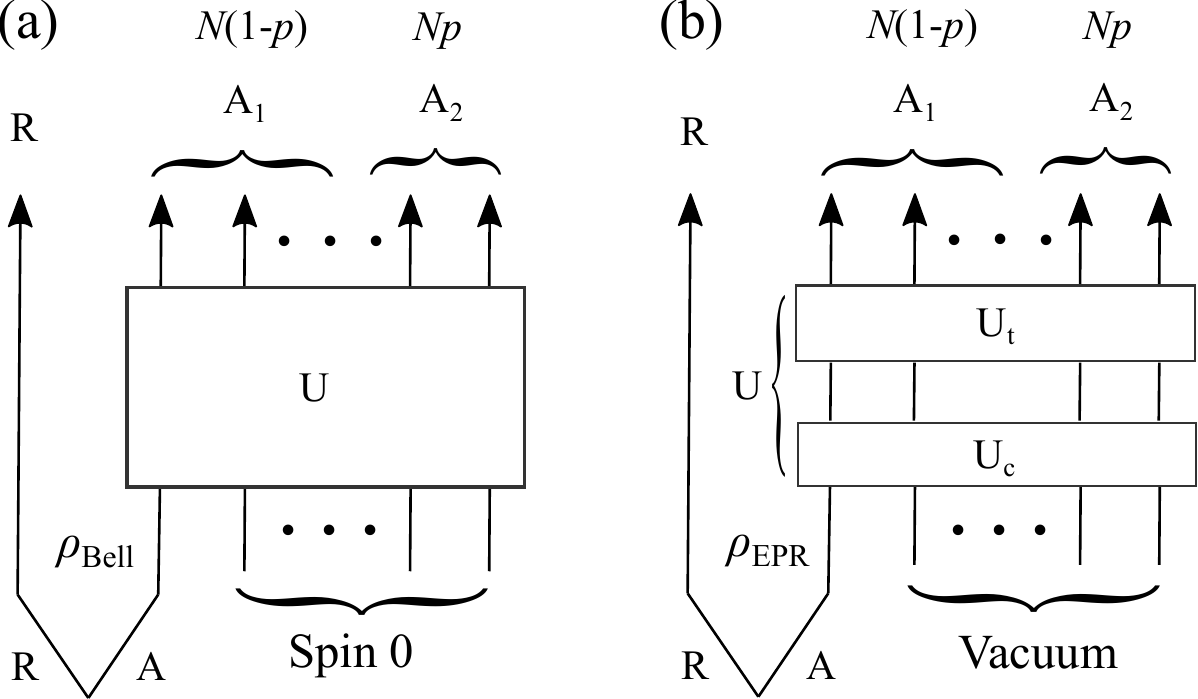}
\caption{The decoupling model of quantum erasure channel for: (a) discrete variable channel where quantum information is scrambled by the Haar random unitary; (b) continuous variable channels where photons are first equally distributed among all the Bosonic modes by the passive random unitary $U_\mathrm{c}$, then further scrambled by the Haar random unitary $U_t$ confining to the effective subspace. \label{fig1}}
\end{figure}

For a DV quantum erasure channel, it is well known that the channel capacity is in principle achievable with a random channel coding \cite{klesse2007}, as depicted in Fig.~\ref{fig1}(a), where the quantum information carried by the input system $A$ is unitarily scrambled with total spins $N$. In the large limit $N$, accessing the $N(1-p)$ spins in the output can recover the scrambled quantum information, respecting the channel capacity of the DV quantum erasure channel. This recovery is indeed possible if the reference system $R$ and the remaining $Np$ qubits are close to a product state, which is the essence of decoupling theory \cite{hayden2008}. {This theory plays a central role in the well-known Hayden-Preskill thought experiment, where one tries to hide information by tossing it into a black hole but unfortunately the information will be quickly reflected out when the black hole is at a certain stage \cite{hayden2007}.}

Naturally, one would consider investigating a similar channel coding model by replacing the spins with Bosonic modes that simulate the CV erasure channel. However, an immediate problem shows up when one notices that the CV system is infinite dimensional. The random coding scheme would involve infinite energy or noncompact group operations \cite{zhuang2019,fukuda2019}, making the further discussion impossible. One might simply think to set a cutoff to get a finite dimensional Hilbert space such that the decoupling reduces to the finite dimensional situation. However, in this case, a maximally mixed state is the best encoding that achieves the optimal information transmission rate $(1-2p)\log_2n_c$ ($n_c$ is the cutoff on the Fock basis) \footnote{See the appendix for a brief review of the calculations for the discrete variable decoupling with any finite dimension.}. With the same amount of energy carried by the maximally mixed state, a thermal state has a larger entanglement entropy, which indicates that it would be better to encode a thermal state in the CV setting without doing any cut-off, as long as we can correctly identify the unitary random scrambling. 

In this paper, we will obtain the proper scrambling unitary which is essential for using the CV decoupling model to understand the CV erasure channel. By defining an effective dimension of a general Bosonic mode, we are able to get the correct Haar random unitary restricted to the effective Hilbert space. Using decoupling formalism in quantum channel theory, we show the information transmission rate with the CV erasure channel is upper bounded by $(1-2p)\mathcal{E}$, where $\mathcal{E}$ represents mode entanglement that is specifically related to the mode's effective dimension. This bound essentially gives the quantum channel capacity since random channel coding is optimal. {Our discussion of the CV erasure channel naturally extends the original Hayden-Preskill thought experiment to the continuous variable case, and probably is more precise in the setting of black hole since the CV model predicts that each radiated mode is thermal, matching the property of Hawking radiation.}

\section{Quantum channel and the decoupling theorem}

Mathematically, a CPTP map that describes a quantum channel $\mathcal{N}:\rho_A\rightarrow\rho_B$
always has a unitary dilation given by \cite{paulsen2002}
\begin{equation}
    \mathcal{N}(\rho_A)\equiv\text{tr}_E[\mathcal{U}_{AE}(\rho_A\otimes\ket{0}\bra{0}_E)]
\end{equation}
where the unitary channel $\mathcal{U}_{AE}$ acts on a dilated Hilbert space $H_{AE}$. By convention,
the subscript $E$ is used to denote the environment degree of freedom. Since unitary evolution correlates the system and the environment, the system output usually cannot recover all the information from the input. However, it is shown that error correction for restoring all information is possible if certain decoupling condition is satisfied \cite{schumacher1996,schumacher2002,hayden2007,hayden2008}.
\begin{theorem}
Denote $\ket{\psi}_{RA}$ as a purification of the input $\rho_A$ ($R$ is generally called the reference system), we have a unitary channel $I_R\otimes\mathcal{U}_{AE}$ acting on the input $\ket{\psi}_{RA}\otimes\ket{0}_E$
\begin{equation}
    \ket{\psi}_{RBE}\bra{\psi}=I_R\otimes\mathcal{U}_{AE}(\ket{\psi}_{RA}\bra{\psi}\otimes\ket{0}_E\bra{0}).
\end{equation}
Perfect entanglement recovery is possible if and only if the reduced density matrix of the reference and the environment are decoupled in the following sense
\begin{equation}
    \rho_{RE}=\rho_R\otimes\rho_E.
\end{equation}
\end{theorem}
This theorem is first given in Refs.~\cite{schumacher1996,schumacher2002} as a condition for perfect quantum error correction. The ``if part" of the theorem is interesting. Intuitively, it means that if the reference system is not correlated with the environment, its entanglement with the system must be preserved in the output. For convenience, we put the ``if part" proof here and more details should resort to Ref.~\cite{schumacher1996}.
\begin{proof}
Suppose that $\rho_R$ and $\rho_E$ have the following spectral decompositions:
\begin{equation}
    \rho_R=\sum_k\lambda_k\ket{k^R}\bra{k^R},\rho_E=\sum_l\mu_l\ket{l^E}\bra{l^E}.
\end{equation}
Then the pure state $\ket{\psi}_{RBE}$ can be written in the form
\begin{equation}
    \ket{\psi}_{RBE}=\sum_{kl}\sqrt{\lambda_k\mu_l}\ket{k^R}\ket{\phi^B_{kl}}\ket{l^E}.
\end{equation}
The states $\ket{\phi^B_{kl}}$ are orthonormal, respecting the Schmidt decomposition. Now we can design a local projection measurement on $B$
\begin{equation}
    \Pi_l=\sum_k\ket{\phi^B_{kl}}\bra{\phi^B_{kl}},
\end{equation}
where each measurement result $l$ occurs with probability $\mu_l$. The post-state conditioning on $l$ is 
\begin{equation}
    \ket{\psi^l_{RBE}}=\sum_k\sqrt{\lambda_k}\ket{k^R}\ket{\phi^B_{kl}}\ket{l^E}.
\end{equation}
Then a unitary transformation $U_l\ket{\phi^B_{kl}}=\ket{k^B}$ yields the state
\begin{equation}
    U_l\ket{\psi^l_{RBE}}=\sum_k\sqrt{\lambda_k}\ket{k^R}\ket{k^B}\otimes\ket{l^E}.
\end{equation}
Obviously, $\ket{\psi_{RB}}=\sum_k\sqrt{\lambda_k}\ket{k^R}\ket{k^B}$ carries the exactly same amount of entanglement as the state $\ket{\psi_{RA}}$.
\end{proof}

For any quantum channel, this decoupling condition is the requirement for perfectly transmitting input states. The decoupling condition is not always satisfied. However, if the condition is approximately true, the entanglement is also approximately recoverable. Formally, it can be rigorously shown in the following theorem:
\begin{theorem}
Let $\mathcal{V}_\mathcal{N}^{A\rightarrow BE}$ be a Stinespring extension of the channel $\mathcal{N}_{A\rightarrow B}$. Let $\rho_{RA^n}$ be any encoding with dimension $\abs{R}=2^{nQ}$. The output state $\rho_{RB^nE^n}=\mathcal{V}_\mathcal{N}^{\otimes n}(\rho_{RA^n})$. Then there exists a decoding map $\mathcal{D}_{B^n\rightarrow \hat{R}}$ which, together with the encoding $\rho_{RA^n}$, comprises a $(Q,n,\epsilon)$ entanglement generation code for the channel, provided that
\begin{equation}
    ||\rho_{RE^n}-\Pi_R\otimes\varphi_{E^n}||_1\le\epsilon,
\end{equation}
where $\Pi_R=\frac{1}{\abs{R}}{I}$.
\end{theorem}
This theorem and proof can be found in Ref.~\cite{hayden2008}. Arbitrary small $\epsilon$ can be achieved for large $n$, indicating that the decoupling between the reference and the environment is asymptotically satisfied as one increases the number of channel usage. Thus it guarantees the existence of decoding operations on the output state to restore almost all the entanglement.

\section{Decoupling model for CV erasure channel}

In the DV decoupling model, generally one party of the entangled qubit (qudit) systems is sent down to a random unitary channel, as shown in Fig.~\ref{fig1}(a). Quantum information is fully scrambled, so that certain local access to the scrambled system cannot fully recover the information, respecting the decoupling theorem \cite{dupuis2010,horodecki2007}. 
Using the decoupling theorem, one can further obtain the expression of quantum capacity for quantum erasure channel \cite{hayden2007,hayden2008,choi2020}. 

\subsection{The CV decoupling model}

We want to extend the discussion to the CV quantum erasure channel. Naively, one would think that we just need to replace those qubits with infinite dimensional modes, e.g., Bosonic modes. For instance, as shown in Fig.~\ref{fig1}(b), assuming two Bosonic modes are in a two-mode squeezed state $\rho_{RA}^\mathrm{EPR}$, one of which ($A$) is sent to a random encoding circuit for information scrambling. The next task then is to find the maximal number of modes that can remain decoupled from the reference mode $R$. As in the DV case, this model is indeed simulating a CV quantum erasure channel with a random coding scheme. However, we need to identify what kind of unitary to be used, which turns out to be non-trivial for CV systems. 

Unlike the DV case, the CV system is infinite dimensional, which might involve unbounded energy \cite{fukuda2019}. The unitary group for fully scrambling the system is generally noncompact \cite{zhuang2019,zhuang2021}, e.g., the Gaussian unitary with squeezing forms a noncompact group. A natural guess is to use random passive unitary $U_\mathrm{c}$ (without squeezing, it forms a compact group) to scramble those Bosonic modes. However, a close examination could quickly give us a negative answer because the passive random unitary does not form a unitary design \footnote{A unitary design is a subset of the unitary group where the sample averages of certain polynomials over the set match that over the whole unitary group.}. Actually, even including the squeezing, the Gaussian unitary is not a perfect unitary design \cite{zhuang2019}, which cannot fully scramble the quantum information.
It is then believed that some nontrivial (certainly non-Gaussian) unitary is required in the CV random coding, and we shall try identifying it in the following sections.

\subsection{The effective dimension of CV system}

In Ref.~\cite{zhuang2019}, many-mode Bosonic system scrambling is discussed in the phase space, where so-called quasi or genuine scrambling is achieved by approximated CV unitary designs. In this paper, we discuss the mode scrambling in a different approach. As mentioned, a Bosonic mode has infinite dimension that could carry infinite energy, while in practice, each mode usually has a finite amount of energy and can be effectively described as a finite dimensional system.
\begin{theorem}
    A single mode Bosonic state $\rho$ with von Neumann entropy $\mathcal{E}=-tr\rho\log\rho$ has an effective dimension
    \begin{equation}
        D=2^\mathcal{E}.
    \end{equation}
\end{theorem}
Note that this theorem just means that the mode can be effective described by a $D$-dimensional qudit in the sense of identifying the entanglement.
\begin{proof}
Expressing the state $\rho$ in its diagonal basis (spectrum decomposition)
\begin{equation}
    \rho=\sum_\lambda p_\lambda\ket{\lambda}\bra{\lambda},
\end{equation}
where $p_\lambda$ is the probability corresponding to $\ket{\lambda}$. Imagine that we have a quantum source emitting this state and we can detect the state in this basis. In general, the emitting state is written as (assuming $K$ copies)
\begin{equation}
    \rho^{\otimes K}=\sum_{\lambda_1,...,\lambda_K}p_{\lambda_1}...p_{\lambda_K}\ket{\lambda_1}\bra{\lambda_1}\otimes...\ket{\lambda_K}\bra{\lambda_K}.
\end{equation}
We would be able to define a typical subspace as $K$ becomes very large. For example, if we measure each mode on the basis, we will get a string of eigenvalues $i_1i_2...i_j...i_K$ with $1\le j\le K,0\le i_j<\infty$. Each string has the probability $p_{i_1}p_{i_2}...p_{i_j}...p_{i_K}$. In those strings, there is a set in which the strings are more probable than the strings outside the set. This set is usually called a typical set, and the probability for each string in the set is given as
\begin{equation}\label{prob1}
    p(\text{typical string})=\prod_{i_j}p_{i_j}^{Kp_{i_j}}.
\end{equation}
 To identify the rank of the typical set, we take the logarithm on both sides of Eq.~\ref{prob1} and obtain
\begin{equation}
\begin{split}
    p(\text{typical string})&=2^{K\sum_{i_j=0}^\infty p_{i_j}\log_2p_{i_j}}\\
    &=2^{-K\mathcal{E}},
\end{split}
\end{equation}
where $\mathcal{E}=-\sum_{i_j=0}^\infty p_{i_j}\log_2p_{i_j}$ is the von Neumann entropy of the state. (Note more rigorously, one should define the typical set by bounding small errors, e.g., $2^{-(K\mathcal{E}+\epsilon)}\le p(\text{typical})\le 2^{-(K\mathcal{E}-\epsilon)}$, and here we take the ideal limit for demonstrating the main ideas, which is good enough for our purpose.) Obviously, each string in the typical set is equally probable, and since the probability sums to one (note that the string outside the typical set almost never happens \cite{shannon1948, wilde2013}), the total number of typical strings is $2^{K\mathcal{E}}$. Alternatively, we can think these $K$ modes span a subspace with dimension $2^{K\mathcal{E}}$, thus each mode effectively contribute to the dimension with  $2^{\mathcal{E}}$. 

\end{proof}

\section{The decoupling and quantum capacity}

As shown in Fig.~\ref{fig1}(b), we consider the model in which a reference $R$ is entangled with the system $A$. Without losing generality, we assume that there are $K$ Bosonic modes in $R$ and $N\gg K$ modes in $A$, and they are initially entangled by $K$ pairs of two-mode squeezed state $\rho_\mathrm{EPR}^{\otimes K}$ (only one pair is shown in the figure). The rest of $N-K$ modes in $A$ are initially set in the vacuum. We then apply a random unitary on the system $A$, hoping to scramble the information in the $N$ modes such that discarding a finite number of modes in $A$ will not decrease the entanglement with $R$. Suppose that for the initial $K$ two-mode squeezed states, each pair carries entanglement $\mathcal{E}(r)$, which is given by
\begin{equation}
    \mathcal{E}(r)\equiv\cosh^2(r)\log_2\cosh^2(r)-\sinh^2(r)\log_2\sinh^2(r),
\end{equation}
where $r$ denotes the strength of the squeeze. Each mode in the two-mode squeezed state contains on average $\sinh^2r$ photons. 

In the system $A$, we first apply passive random unitary $U_\mathrm{c}$ (random beam splitter and phase shifter) to redistribute $K\sinh^2r$ photons to totally $N$ modes, and {on average each mode has} $K\sinh^2r/N$ photons. Note the passive random unitary on average does not thermalize the state in the sense (see the appendix for more details)
\begin{equation}
    \rho_{RA}^\prime\equiv\int dU_c U_c\rho_{RA}U_c^\dagger\neq\rho_R\otimes\rho_A^\mathrm{thermal}.
\end{equation}
Instead, we have $\rho^\prime_{RA}=\rho_R\otimes(\tilde{\rho}^{\otimes N})_A$ where the state $\tilde{\rho}$ for each different mode is exactly the same and is diagonal on the Fock basis
\begin{equation}
   \tilde{\rho}=\sum_n p_n\ket{n}
   \bra{n}
\end{equation}
since the random phase shifter washes out all correlations in that basis. The entanglement of each mode is given by $\mathcal{E}(p_n)=-\sum_n p_n\log_2 p_n$. Accordingly, the typical subspace of $A$ has a finite dimension
\begin{equation}
    d=2^{N\mathcal{E}(p_n)}
\end{equation}
which can effectively describe the infinite dimensional Bosonic modes. Next, we apply a second random unitary $U_t$ on the system $A$ that is confined to this typical subspace, and we shall show below that the decoupling inequality can be recovered. Note that according to the equipartition theorem in the large mode limit \cite{wilde2013}, the typical projection measurement will have an approximately uniform distributed outcome. The second random unitary $U_t$ essentially gives an uniform distribution of the typical state, and each mode is close to a thermal state with photon number $\sinh^2r_0=K\sinh^2r/N$. The corresponding entanglement is denoted $\mathcal{E}(r_0)$, as detailed in the Appendix.

Now, we divide the system $A$ into two parts $A_1$ and $A_2$. $A_2$ contains $pN$ modes ($0<p<1$), which is to be discarded. The question is how many modes in $A$ can we discard while keeping the remaining modes almost completely entangled with the reference $R$. 
{The answer is summarized in the following theorem:
\begin{theorem}\label{theo4}
The average trace distance between the randomized state on $RA_2$ and a product state $\rho_R\otimes\prod_{A_2}$ satisfies the following inequality
\begin{equation}\label{decoi}
     \int dU||\text{tr}_{A_1}[U\rho_{RA}U^\dagger]-\rho_R\otimes\Pi_{A_2}||_1\le 2^{-\frac{1}{2}(Q-\gamma)N},    
\end{equation}
where the unitary $U=U_tU_c$ acts on the system $A$. $Q=(1-2p)\mathcal{E}(r_0)$ is the energy-constrained quantum capacity and $\gamma=K\mathcal{E}(r)/N$ is the quantum information transmission rate. The product state $\rho_R\otimes\Pi_{A_2}$ is defined as the unitary average
\begin{equation}\label{ruave}
    \rho_R\otimes\Pi_{A_2}\equiv \text{tr}_{A_1}\int dU_t U_t\rho^\prime_{RA}U_t^\dagger.
\end{equation}
\end{theorem}
In the language of the decoupling principle, to recover all initial entanglement from the system $RA_1$, we need the term $2^{-\frac{1}{2}(Q-\gamma)N}$ to be small, which is achievable in the large $N$ limit if $\gamma<Q$. This clearly indicates that all the information transmission rate $\gamma$ is achievable as long as it is smaller than the quantum channel capacity.}

To prove the inequality, we start from the following equality: for arbitrary state $\rho$, the average over Haar random unitary confined to the typical subspace satisfies
\begin{equation}
    \int dU_t U_t\rho U_t^\dagger=\frac{1}{d}I.
\end{equation}
where $d$ is the typical space dimension defined above and $I$ is the corresponding identity operator. Intuitively, it is because the integral must commute with any unitary operation (the property of Haar measure), thus only identity is possible. Rigorously, we have
\begin{align}
    \int dU_t U_t\rho U_t^\dagger=&\int dU_t\sum_{i_1j_1,i_1^\prime j_1^\prime} U^t_{i_1j_1}\rho_{j_1i_1^\prime}{U^t}^\dagger_{i_1^\prime j_1^\prime}\\
    =&\sum_{i_1j_1,i_1^\prime j_1^\prime}\rho_{j_1i_1^\prime}\int dU_t U^t_{i_1j_1}{U^t}^\dagger_{i_1^\prime j_1^\prime}\\
    =&\sum_{i_1j_1,i_1^\prime j_1^\prime}\rho_{j_1i_1^\prime}\frac{1}{d}\delta_{i_1j_1^\prime}\delta_{j_1i_1^\prime}\\
    =&\sum_{i_1j_1}\rho_{j_1j_1}\frac{1}{d}\delta_{i_1 i_1}=\sum_{i_1}\frac{\sum_{j_1}\rho_{j_1j_1}}{d}\delta_{i_1i_1}\\
    =&\frac{1}{d}\sum_{i_1}\delta_{i_1i_1}\ket{i_1}\bra{i_1}\label{basisre},
\end{align}
where the basis is omitted from the above calculation for simplicity and is recovered in the last line~\ref{basisre}. 
Note in Eq.~\ref{ruave}, the random unitary is acting on the system $A$, and we expect on average the state to be
\begin{equation}
\begin{split}
    &\int dU U\rho_{RA}U^\dagger\\
    =&{\int dU_t U_t\left(\int dU_c U_c\rho_{RA}U_c^\dagger\right) U_t^\dagger}\\
    =&\rho_R\otimes\frac{1}{d}{I}_A=\rho_R\otimes\Pi_A,
\end{split}    
\end{equation}
where $\Pi_{A}$ is to denote a maximally mixed state in the typical subspace. Obviously, it is also true if we confine the system to $A_2$
\begin{equation}
     \text{tr}_{A_1}\int dU U\rho_{RA}U^\dagger=\rho_R\otimes\frac{1}{d}I_{A_2}=\rho_R\otimes\Pi_{A_2}.
\end{equation}
which is the expression defined in Eq.~\ref{ruave}. This result basically means that the correlation between $R$ and $A$ will be averaged out by the random unitary, and the information is completely scrambled in the sense that $A$ is set in a maximally mixed state. Note that for each unitary realization, it is not necessarily true. 
In the following,  we want to show that for each unitary realization, the distance between $\text{tr}_{A_1}[U\rho_{RA}U^\dagger]$ and $\rho_R\otimes\Pi_{A_2}$ is typically small in the limit of large $N$ Bosonic modes. This is guaranteed if we can somehow bound the variance as in Eq.~\ref{decoi} (the average trace distance) as follows
\begin{widetext}
\begin{align}
    &\int dU||\text{tr}_{A_1}[U\rho_{RA}U^\dagger]-\rho_R\otimes\Pi_{A_2}||_1\\
    \le&\int dU\sqrt{d_{RA_2}}||\text{tr}_{A_1}[U\rho_{RA}U^\dagger]-\rho_R\otimes\Pi_{A_2}||_2 \label{eqabc1}\\
   =&\int dU\sqrt{d_{RA_2}}\sqrt{\text{tr}_{RA_2}(\text{tr}_{A_1}[U\rho_{RA}U^\dagger]-\rho_R\otimes\Pi_{A_2})^2}\\
   \le&\sqrt{d_{RA_2}}\sqrt{\text{tr}_{RA_2}\int dU (\text{tr}_{A_1}[U\rho_{RA}U^\dagger]-\rho_R\otimes\Pi_{A_2})^2} \label{eqabc2}\\
   =&\sqrt{d_{RA_2}\text{tr}_{RA_2}\int dU\left[ (\text{tr}_{A_1}U\rho_{RA}U^\dagger )^2-(\rho_R\otimes\Pi_{A_2})^2  \right] } \\
   =&\sqrt{d_{RA_2}\left(\text{tr}_{RA_2}\int dU (\text{tr}_{A_1}U\rho_{RA}U^\dagger )^2-\text{tr}_{RA_2}(\rho_R\otimes\Pi_{A_2})^2  \right)} \label{eqabc}.
\end{align}
\end{widetext}
The line \ref{eqabc1} is obtained since the trace norm is upper bounded by the Hilbert-Schmidt norm up to the dimension factor $d_{RA_2}$. The line \ref{eqabc2} is due to the Jensen's inequality.
Obviously, we only need to evaluate one term involving the random unitary, which is
\begin{widetext}
\begin{align}
    &\text{tr}_{RA_2}\int dU(\text{tr}_{A_1}U\rho_{RA}U^\dagger)^2\\
    =&\int dU \text{tr}_{RAR^\prime A^\prime}\left[\rho_{RA}(U)\otimes\rho_{R^\prime A^\prime}(U)F_{RR^\prime} F_{A_2A_2^\prime}I_{A_1A_1^\prime}\right] \label{eqswap} \\
    =&\int dU_c \text{tr}_{RAR^\prime A^\prime}\left[\rho_{RA}(U_c)\otimes\rho_{R^\prime A^\prime}(U_c) F_{RR^\prime}\int dU_t {U_t^\dagger}^{\otimes 2}  F_{A_2A_2^\prime}I_{A_1A_1^\prime}U_t^{\otimes 2}\right],\label{eqab}
\end{align}
\end{widetext}
where the notation $\rho(U)=U\rho U^\dagger$ is used and we use $F$ to denote a swap operator. The line \ref{eqswap} is due to the swap technique
\begin{equation} \text{tr}_A(\sigma_A^2)=\text{tr}_{AA^\prime BB^\prime}[\sigma_{AB}\otimes\sigma_{A^\prime B^\prime}F_{AA^\prime}I_{BB^\prime}].
\end{equation}
We see the last integral $\int dU U^\dagger\otimes U^\dagger F_{A_2A_2^\prime}I_{A_1A_1^\prime}U\otimes U$ is all we need that involves random unitary in the typical space. Now we use the double-twirling formula \cite{collins2006} (which can be directly obtained from the Weingarten calculus)
\begin{widetext}
\begin{equation}
\begin{split}
    &\int dU_t U_t^\dagger\otimes U_t^\dagger X U_t\otimes U_t \\
    =&  \left[\text{tr}(X)-\frac{\text{tr}(XF_{AA^\prime})}{d}\right]\frac{I_{AA^\prime}}{{d_A^2-1}}+\left[\text{tr}(XF_{AA^\prime})-\frac{\text{tr}(X)}{d}\right]\frac{F_{AA^\prime}}{{d_A^2-1}} \\
    =&\frac{1}{d_{A_2}}\frac{1-1/d_{A_1}^2}{1-1/d_A^2}I_{AA^\prime} + \frac{1}{d_{A_1}}\frac{1-1/d_{A_2}^2}{1-1/d_A^2}F_{AA^\prime}\label{eqa}\\
    \le&\frac{1}{d_{A_2}}I_{AA^\prime}+\frac{1}{d_{A_1}}F_{AA^\prime}
\end{split}
\end{equation}
\end{widetext}
where we take $X=F_{A_2A_2^\prime}\otimes I_{A_1A_1^\prime}$. In the last equality, we used the results
\begin{align}
    \text{tr}(X) = d_{A_1}^2d_{A_2},
    \text{tr}(XF_{AA^\prime}) = d_{A_1}d_{A_2}^2.
\end{align}
The last inequality is obtained by considering the fact that the typical dimensions for the system $A$ and subsystems $A_{1,2}$  are much larger than one. Now we can continue to evaluate Eq.~\ref{eqab}, which is
\begin{widetext}
\begin{align}
     &\int dU_c \text{tr}_{RAR^\prime A^\prime}\left[\rho_{RA}(U_c)\otimes\rho_{R^\prime A^\prime}(U_c) F_{RR^\prime}\int dU_t {U_t^\dagger}^{\otimes 2} F_{A_2A_2^\prime}I_{A_1A_1^\prime}U_t^{\otimes 2}\right]\\
     \le&\int dU_c \text{tr}_{RAR^\prime A^\prime}\left[\rho_{RA}(U_c)\otimes\rho_{R^\prime A^\prime}(U_c) F_{RR^\prime}(\frac{1}{d_{A_2}}I_{AA^\prime}+\frac{1}{d_{A_1}}F_{AA^\prime}) \right]\\
     =&\frac{1}{d_{A_2}}\text{tr}_{R}\rho_{R}^2+\frac{1}{d_{A_1}}\text{tr}_{AR}\rho_{AR}^2.
\end{align}
\end{widetext}
Then we are ready for the calculation of Eq.~\ref{eqabc} using the integral we just derived
\begin{equation}\begin{split}
    &\sqrt{d_{RA_2}}\sqrt{\text{tr}_{RA_2}\int dU (\text{tr}_{A_1}U\rho_{RA}U^\dagger )^2-\text{tr}_{RA_2}(\rho_R\otimes\Pi_{A_2})^2  }\\
    \le&\sqrt{d_{RA_2}}\sqrt{\frac{1}{d_{A_2}}\text{tr}_{R}\rho_{R}^2+\frac{1}{d_{A_1}}\text{tr}_{AR}\rho_{AR}^2-\text{tr}_{RA_2}(\rho_R\otimes\Pi_{A_2})^2  }\\
    =&\sqrt{\frac{d_{RA_2}}{d_{A_1} }\text{tr}_{AR}\rho_{AR}^2 },
\end{split}\end{equation}
where the line is obtained since $\text{tr}_{RA_2}(\rho_R\otimes\Pi_{A_2})^2=\text{tr}_R\rho_R^2/d_{A_2}$. To summarize, we finally obtain the upper bound
\begin{equation}
    \int dU||\text{tr}_{A_1}[U\rho_{RA}U^\dagger]-\rho_R\otimes\Pi_{A_2}||_1\le \sqrt{\frac{d_{RA_2}}{d_{A_1} }\text{tr}_{AR}\rho_{AR}^2 },
\end{equation}
which is the main result we have for the CV system using effective dimension. Since we are starting from a pure state $\rho_{AR}$, we have $tr\rho_{AR}^2=1$. Also the typical space dimension of $A_1$ and $RA_2$ are given by
\begin{equation}
    d_{A_1}=2^{(1-p)N\mathcal{E}(r_0)}, d_{RA_2}=2^{K\mathcal{E}(r)+pN\mathcal{E}(r_0)}.
\end{equation}
Thus the upper bound is evaluated to be
\begin{align}
     &\int dU||\text{tr}_{A_1}[U\rho_{RA}U^\dagger]-\rho_R\otimes\Pi_{A_2}||_1\\
     \le&\sqrt{\frac{d_{RA_2}}{d_{A_1}}}
     =2^{\frac{K}{2}\mathcal{E}(r)-(\frac{1}{2}-p)N\mathcal{E}(r_0)}.
\end{align}
The above inequality can be understood in the quantum channel language: we initially have $K\mathcal{E}(r)$ ebits entanglement and try to send the entanglement through $N$ quantum erasure channels with erasure probability $p$. Obviously, the entanglement transmission rate is $\gamma\equiv K\mathcal{E}(r)/N$. Then the upper bound can be rewritten in the form
\begin{equation}
     \int dU||\text{tr}_{A_1}[U\rho_{RA}U^\dagger]-\rho_R\otimes\Pi_{A_2}||_1\le2^{-\left[(\frac{1}{2}-p)\mathcal{E}(r_0)-\frac{\gamma}{2}\right]N},    
\end{equation}
which is the main result of this paper. {The proof to the inequality Eq.~\ref{decoi} is now complete.} 

We see if $(1/2-p)\mathcal{E}(r_0)-\gamma/2>0$, the bound goes to zero in the $N\rightarrow\infty$ limit. According to the decoupling theorem, the entanglement can be recovered by decoding $A_1$ ($A_2$ discarded), which means that any entanglement transmission rate satisfying $\gamma<(1-2p)\mathcal{E}(r_0)$ is asymptotically achievable. Since $\mathcal{E}(r_0)$ essentially constrains the input energy for each mode, we define
\begin{equation}
    Q\equiv(1-2p)\mathcal{E}(r_0)
\end{equation}
as the \textit{energy-constrained} quantum channel capacity for continuous variable quantum erasure channel using the random coding scheme.
Interestingly, $1-2p$ is actually the quantum capacity of qubit erasure channel with erasure probability $p$, which quantitatively matches the special case when $\mathcal{E}(r_0)=1$.

\section{Discussion and outlook}

\begin{figure}[t]
\centering
\includegraphics[width=\columnwidth]{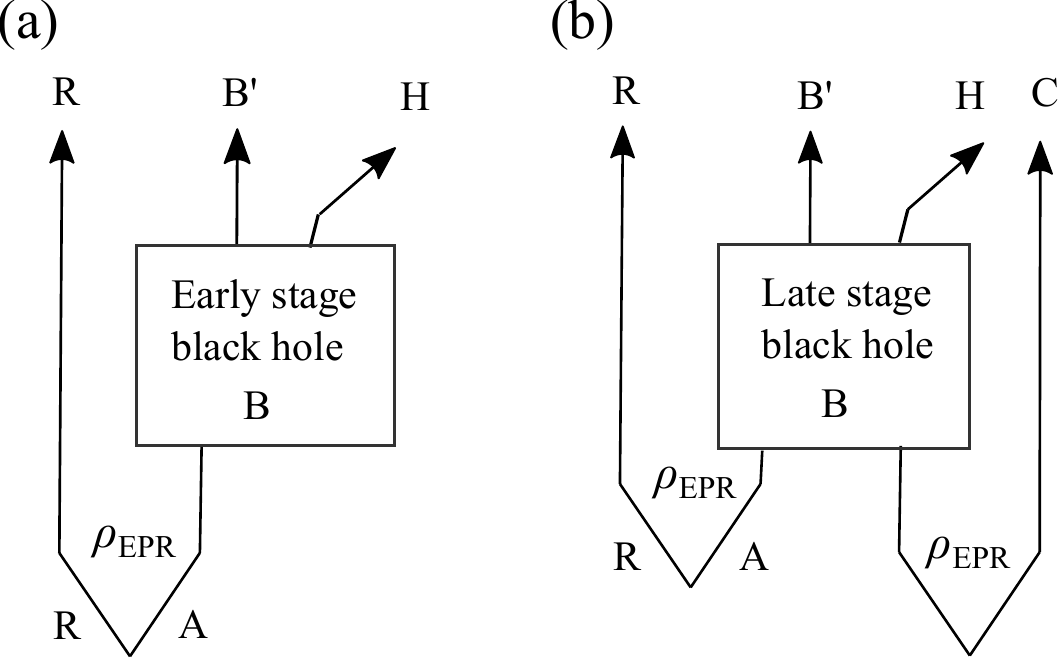}
\caption{{Black holes as mirrors: (a) Alice's information (A) is tossed into an early stage black hole (B). Her information will be revealed as soon as the Hawking radiation (H) accumulates half the degrees of freedom of the black hole; (b) Alice's information is tossed into a late-stage black hole, where Charlie (C) has already collected a certain amount of Hawking radiation thus is entangled with the black hole. Her information will quickly revealed to Charlie when a little bit more black hole radiation is collected.} \label{fig2}}
\end{figure}

{The DV erasure channel was used in the Hayden-Preskill thought experiment \cite{hayden2007}, where black holes turn out to be a bad keeper of secret information. As shown in Fig.~\ref{fig2}, Alice (A) tries to hide her secret in a black hole, assuming the black hole has a fast enough dynamics to scramble the information. Quite in contrast, if an eavesdropper collects the Hawking radiation to a certain degree, i.e., a little bit more than half of the black hole degrees of freedom, he could know Alice's secret accurately. If the black hole was in its late stage, i.e., half of the black hole was evaporated, and the eavesdropper has monitored the black hole dynamics and collected all the Hawking radiation, then Alice's secret will be quickly revealed to the eavesdropper as long as a little bit more Hawking radiation is obtained. The reason is deep rooted in the quantum capacity of the DV erasure channel. With the Theorem~\ref{theo4} that we prove in this paper, this thought experiment naturally extends to the CV setting. More interestingly, the CV model might be more proper in describing the black hole dynamics, since the Hawking radiation is believed to be thermal. Because each Bosonic mode is in a thermal state after random scrambling, it thus correctly models the black hole radiation.}

On a glance of the literature, it is easy to find that the DV-system-based theory of quantum information and computation is more advanced, while the CV counterpart is less developed. Admittedly, the recent decade sees the gradual progress of CV quantum information processing \cite{weedbrook2012,braunstein2005}, e.g., encoding and error correcting quantum information with harmonic oscillators \cite{victor2018,ma2021}. Although it is still hard to get the full controllability of the CV system, its infinite dimension with potential to encode larger amounts of quantum information is appealing and is usually called hardware efficient for a single system. Thus identifying the error model of CV channels and the corresponding correcting schemes are essential.

As well known, evaluating the exact capacity of quantum channels is notoriously difficult that involves optimizing two-letter functions \cite{gyongyosi2018,bradler2015}. The decoupling theorem looks into the capacity in a different way, directly calculating a random coding scheme, which is used a lot in discrete variable quantum channels. By confining the random unitary on the typical subspace of the CV systems, we derived the capacity of the CV quantum erasure channel, which avoids the problem of non-compact unitary scrambling. We expect the technique developed in the paper to shed new light on exploring other types of CV quantum channel, e.g., the long-standing capacity bound problem of the well-known Bosonic loss channels, and we leave that for future investigations.

\begin{acknowledgments}
We thank Junyu Liu for helpful discussions. We acknowledge supports from the ARO (W911NF-18-1-0020, W911NF-18-1-0212), ARO MURI (W911NF-16-1-0349, W911NF-21-1-0325), AFOSR MURI (FA9550-19-1-0399, FA9550-21-1-0209), AFRL (FA8649-21-P-0781), DoE Q-NEXT, NSF (OMA-1936118, EEC-1941583, OMA-2137642), NTT Research, and the Packard Foundation (2020-71479).

\end{acknowledgments}

\bibliography{all}
\onecolumngrid
\begin{appendix}

\section{The Weingarten calculus}

The Weingarten calculus is important in random matrix theory, in which an integral is repeatedly used
\begin{equation}\label{eqapp1}
    \int dU U_{i_1j_1}U_{i_2j_2}...U^\dagger_{j_1^\prime i_1^\prime}U^\dagger_{j_2^\prime i_2^\prime}...=\sum_{\sigma,\tau}\delta_{i_1i_{\sigma(1)}^\prime}\delta_{i_2i_{\sigma(2)}^\prime}...\delta_{j_1j_{\tau(1)}^\prime}\delta_{j_2j_{\tau(2)}^\prime}...W_g(\sigma^{-1}\tau),
\end{equation}
where $W_g$ is called Weingarten function that depends on the permutations $\sigma$ and $\tau$. In the main text, we already use this formula to derive $\int dU U\rho U^\dagger=I/d$. Another formula that appears frequently in decoupling theory is the double unitary twirling formula
\begin{equation}\begin{split}
    \mathcal{T}(X)=&\int dU U^\dagger\otimes U^\dagger X U\otimes U\\
    =&\frac{I}{d^2-1}\left[trX-\frac{Tr(XF)}{d} \right]+\frac{F}{d^2-1}\left[tr(XF)-\frac{trX}{d} \right]
\end{split}\label{eqapp2}
\end{equation}
In many literature, this result is simply stated as the consequence of Schur-Weyl duality from group representation theory, which is not friendly to beginners. The Schur-Weyl duality might be needed to derive Eq.~\ref{eqapp1}, but the double twirling formula Eq.~\ref{eqapp2} can be derived straightforwardly by applying Eq.~\ref{eqapp1}, as shown below:
\begin{equation}
\begin{split}
    \mathcal{T}(X)=&\int dU U^\dagger\otimes U^\dagger X U\otimes U\\
    =&\int dU\sum_{\substack{i_1j_1i_2j_2 \\ i_1^\prime j_1^\prime i_2^\prime j_2^\prime}} U_{i_1j_1}U_{i_2j_2}X^{j_1j_2}_{i_1^\prime i_2^\prime}U_{i_1^\prime j_1^\prime}^\dagger U_{i_2^\prime j_2^\prime}^\dagger \\
    =&\sum_{\substack{i_1j_1i_2j_2 \\ i_1^\prime j_1^\prime i_2^\prime j_2^\prime}}X^{j_1j_2}_{i_1^\prime i_2^\prime}\int dU U_{i_1j_1}U_{i_2j_2}U_{i_1^\prime j_1^\prime}^\dagger U_{i_2^\prime j_2^\prime}^\dagger\\
    =&\sum_{\substack{i_1j_1i_2j_2 \\ i_1^\prime j_1^\prime i_2^\prime j_2^\prime}}X^{j_1j_2}_{i_1^\prime i_2^\prime}\sum_{\sigma,\tau}\delta_{i_1j^\prime_{\sigma(1)}}\delta_{i_2j^\prime_{\sigma(2)}}\delta_{j_1i^\prime_{\tau(1)}}\delta_{j_2i^\prime_{\tau(2)}}W_g(\sigma^{-1}\tau)\\
    =&\sum_{\substack{i_1j_1i_2j_2 \\ i_1^\prime j_1^\prime i_2^\prime j_2^\prime}}X^{j_1j_2}_{i_1^\prime i_2^\prime}[\delta_{i_1j^\prime_{1}}\delta_{i_2j^\prime_{2}}\delta_{j_1i^\prime_{1}}\delta_{j_2i^\prime_{2}}W_g(\sigma_1\tau_1)+ \delta_{i_1j^\prime_{1}}\delta_{i_2j^\prime_{2}}\delta_{j_1i^\prime_{2}}\delta_{j_2i^\prime_{1}}W_g(\sigma_1\tau_2)\\ &\quad\quad\quad\quad+\delta_{i_1j^\prime_{2}}\delta_{i_2j^\prime_{1}}\delta_{j_1i^\prime_{1}}\delta_{j_2i^\prime_{2}}W_g(\sigma_2\tau_1) +\delta_{i_1j^\prime_{2}}\delta_{i_2j^\prime_{1}}\delta_{j_1i^\prime_{2}}\delta_{j_2i^\prime_{1}}W_g(\sigma_2\tau_2)]\\
    =&trX\frac{I}{d^2-1}+tr(XF)\frac{-I}{d(d^2-1)}+trX\frac{-F}{d(d^2-1)}+tr(XF)\frac{F}{(d^2-1)}\\
    =&\frac{I}{d^2-1}\left[trX-\frac{tr(XF)}{d}\right]+\frac{F}{d^2-1}\left[tr(XF)-\frac{trX}{d}\right].
\end{split}
\end{equation}
Note for simplicity, we left out the basis during the calculation and recover that in the end. $F$ is the swap operator acting on the tensor product space
\begin{equation}
    F=\sum_{i_1,j_1^\prime,i_2,j_2^\prime}\ket{i_1}\bra{j_1^\prime}\otimes\ket{i_2}\bra{j_2^\prime}\delta_{i_1j^\prime_2}\delta_{i_2j_1^\prime}.
\end{equation}
Note the permutation group on a string with length two has only two elements $\{\sigma_1(1\rightarrow 1,2\rightarrow 2),\sigma_2(1\rightarrow 2,2\rightarrow 1)\}$, and similar for $\tau$. We also used the Weingarten formula
\begin{align}
    &W_g(\sigma_1\tau_1)=W_g(\sigma_2\tau_2)=\frac{1}{d^2-1},\\
   & W_g(\sigma_1\tau_2)=W_g(\sigma_2\tau_1)=\frac{-1}{d(d^2-1)}.
\end{align}

\section{Random passive operations does not thermalize}
Let us consider a state
\begin{align}
    \rho_\text{in}=\rho_\text{T}(\bar{n})\otimes |0\rangle\langle 0|\otimes |0\rangle\langle 0|\otimes \dots |0\rangle\langle 0|\otimes |0\rangle\langle 0|,
\end{align}
and we apply random passive unitary operations to the state, where $\rho_\text{T}(\bar{n})$ is a thermal state with mean photon number $\bar{n}$.
We claim that after random passive unitary operations, it does not transform to a product of thermal states,
\begin{align}
    \int dU_c U_c\rho_\text{in}U_c^\dagger \neq \rho_\text{T}(\bar{n}/N)\otimes \dots \otimes \rho_\text{T}(\bar{n}/N),
\end{align}
{where $N$ is the number of Bosonic modes}. To see this, let us compare the state projected on the single-photon subspace.
First, the lhs is
\begin{equation}\begin{split}
    &\int dU_c U_cp_1(|1\rangle\langle 1|\otimes |0\rangle\langle 0|\otimes \dots \otimes |0\rangle\langle0|)U_c^\dagger \\ 
    &=\frac{p_1}{N}(|1\rangle\langle 1|\otimes |0\rangle\langle 0|\otimes \dots \otimes |0\rangle\langle0|+|0\rangle\langle 0|\otimes |1\rangle\langle 1|\otimes \dots \otimes |0\rangle\langle0|+\dots+|0\rangle\langle 0|\otimes |0\rangle\langle 0|\otimes \dots \otimes |1\rangle\langle 1|),
\end{split}\end{equation}
where $p_1=\frac{\bar{n}}{(\bar{n}+1)^2}$.
On the other hand, the rhs is
\begin{align}
    q_1(|1\rangle\langle 1|\otimes |0\rangle\langle 0|\otimes \dots \otimes |0\rangle\langle0|+|0\rangle\langle 0|\otimes |1\rangle\langle 1|\otimes \dots \otimes |0\rangle\langle0|+\dots+|0\rangle\langle 0|\otimes |0\rangle\langle 0|\otimes \dots \otimes |1\rangle\langle 1|),
\end{align}
where $q_1=\frac{\bar{n}/N}{(\bar{n}/N+1)^2}$.
Since $p_1/N\neq q_1$, they are not equal.

In the limit of large number of modes, the state of $N_1$ ($N_1\ll N_2\equiv N-N_1$) modes after random average will be typically close to the thermal state in the following sense
\begin{equation}
    \int dU_c||\text{tr}_{N_2}[U_c\rho_\mathrm{in}U_c^\dagger]-\rho^{\otimes N_1}_\mathrm{T}||_1\ll\epsilon,
\end{equation}
which can be straightforwardly shown in the phase space and we refer the reader to the nice paper Ref.~\cite{fukuda2019} for more details. 

Each mode will be in diagonal form in the Fock basis, since the random phase shift in each mode gives
\begin{equation}
\begin{split}
    &\frac{1}{2\pi}\int d\theta e^{i\theta\hat{a}^\dagger\hat{a}}\rho e^{-i\theta\hat{a}^\dagger\hat{a}}\\
    =&\frac{1}{2\pi}\int d\theta e^{i\theta\hat{a}^\dagger\hat{a}}\sum_{m,n}\rho_{mn}\ket{m}\bra{n} e^{-i\theta\hat{a}^\dagger\hat{a}}\\
    =&\frac{1}{2\pi}\sum_{m,n}\int d\theta e^{i\theta (m-n)}\rho_{mn}\ket{m}\bra{n}\\
    =&\sum_{n}\rho_{nn}\ket{n}\bra{n}.
\end{split}    
\end{equation}

\section{Decoupling with truncated Fock space doesn't recover the CV erasure channel capacity}

The typical subspace in the main text gives an efficient way of avoiding the infinite dimension problem in the CV random scrambling unitary. Naively, we might think we can achieve the same goal by simply truncating each mode in the Fock basis. In this appendix, we show this is not true.

Denote in each mode the truncation as $n_c$ in the Fock basis. Thus, the EPR state can be approximated as
\begin{equation}
    \rho_{EPR}^{n_c}=\sum_{n=0}^{n_c}\frac{\tanh^{n}(r)}{\cosh(r)}\ket{n}\ket{n}.
\end{equation}
The entanglement entropy of this state is denoted as $\mathcal{E}_c(r)$ which is in general smaller than $\mathcal{E}(r)$ (the true entanglement of the EPR state), but they will be very close as long as the cutoff number $n_c$ is chosen to be large enough. Now in the same notation, we want to show the decoupling inequality in the discrete variable setting in the hope of revealing the CV erasure quantum capacity
\begin{equation}
\int dU||\text{tr}_{A_1}({U}\rho_{RA}U^\dagger)-\rho_R\otimes \Pi_{A_2}||_1\le \epsilon.
\end{equation}
Note each of the $N$ mode in system $A$ is truncated to have dimension $n_c$ and the random unitary $U$ is acting on a Hilbert space with dimension $n_c^N$. Within this finite dimensional Hilbert space, we have
\begin{align}
    &\int dU||\text{tr}_{A_1}({U}\rho_{RA}U^\dagger)-\rho_R\otimes \Pi_{A_2}||_1\\
    \le&\int dU\sqrt{d_{RA_2}}||\text{tr}_{A_1}({U}\rho_{RA}U^\dagger)-\rho_R\otimes \Pi_{A_2}||_2\\
    =&\sqrt{d_{RA_2}}\int dU\sqrt{ \mathrm{tr}_{RA_2}[(\text{tr}_{A_1}({U}\rho_{RA}U^\dagger)-\rho_R\otimes \Pi_{A_2})^2]}\\
    \le&\sqrt{d_{RA_2}}\sqrt{\int dU\mathrm{tr}_{RA_2}\left[ (\text{tr}_{A_1}[{U}\rho_{RA}U^\dagger])^2]-\mathrm{tr}_{RA_2}[(\rho_R\otimes \Pi_{A_2})^2\right]}\\
    \le&\sqrt{d_{RA_2}}\sqrt{\frac{1}{d_{A_2}}\text{tr}_{R}\rho_{R}^2+\frac{1}{d_{A_1}}\text{tr}_{RA}\rho_{RA}^2 -\mathrm{tr}_{RA_2}[(\rho_R\otimes \Pi_{A_2})^2]}\label{ec7} \\
    =&\sqrt{d_{RA_2}}\sqrt{\frac{1}{d_{A_1}}\text{tr}_{RA}\rho_{RA}^2},
\end{align}
where we used the result $\int dU\mathrm{tr}_{RA_2}[(\text{tr}_{A_1}[U\rho_{RA}U^\dagger])^2] \le\frac{1}{d_{A_2}}\text{tr}_{R}\rho_{R}^2+\frac{1}{d_{A_1}}\text{tr}_{AR}\rho_{AR}^2$ in line \ref{ec7}. Taking the pure initial condition $\text{tr}_{RA}\rho_{RA}^2=1$, we have
\begin{equation}
    \int dU||\text{tr}_{A_1}({U}\rho_{RA}U^\dagger)-\rho_R\otimes \Pi_{A_2}||_1\le\sqrt{ \frac{d_{RA_2}}{d_{A_1}} }.
\end{equation}
The dimension factors on the rhs is determined by the truncation, which has nothing to do with the encoded CV state entanglement. While in the main text, the effective dimension is directly related to the encoded CV state entanglement, enabling us to obtain the CV erasure channel capacity. Actually, the rhs gives 
\begin{equation}
    \sqrt{ \frac{d_{RA_2}}{d_{A_1}} }=2^{-N(Q-\gamma)},
\end{equation}
where $\gamma=\frac{K}{N}\log_2n_c$ and $Q=(1-2p)\log_2n_c$, indicating the $n_c$-dimensional Bell state is the optimal encoding in the decoupling scheme. Imposing energy constraint in the encoded state, we know thermal state would has the maximal entanglement entropy, which justify the CV decoupling scheme shown in the main text is better.

\section{Each mode is a thermal state on average}

As stated in the main text, after the random unitary each mode of $A$ is on average in a thermal state with thermal photon $K\sinh^2{r}/N$. We have
\begin{equation}\begin{split}
    &\int dU U\rho_{RA}U^\dagger
    ={\int dU_t U_t\left(\int dU_c U_c\rho_{RA}U_c^\dagger\right) U_t^\dagger}
    =\rho_R\otimes\frac{1}{d}{I}_A\equiv\rho_R\otimes\rho^\prime_A.
    \end{split}
\end{equation}
$\rho^\prime_A=\frac{1}{d}I_A$ describes the state of $A$ after the random unitary, which essentially is a maximally mixed state in the typical subspace. Note each typical vector in the typical subspace represents a state with an average photon number $K\sinh^2(r)$. Thus on average the state $\rho^\prime_A$ has a total photon number $K\sinh^2(r)$ and each mode in $A$ has $K\sinh^2(r)/N$ photons. 

Now we will further show the reduced density operator for each single mode in $A$ is in a thermal state. Since $\rho^\prime_A$ is a maximally mixed state in the typical subspace, each typical state is equally probable, which essentially defines a micro-canonical ensemble with fixed energy, thus any small subsystem will looks like thermal. To show that, we express the state as
\begin{equation}
\begin{split}
    \rho^\prime_A=&\sum_n P_n\ket{\Phi_n}\bra{\Phi_n}\\
    =&\sum_nP_n\sum_{i\alpha,j\beta}C^n_{i\alpha}C^{n*}_{j\beta}\ket{\phi_i\psi_\alpha}\bra{{\phi_j\psi_\beta}},
\end{split}
\end{equation}
where $\ket{\Phi_n}$ denotes the typical state (energy eigen-states) and $P_n=1/d$ is the probability (density of state independent of $n$). In order to get the reduced density matrix of a single mode, we separate it from the rest $N-1$ modes, and we take $\ket{\Phi_n}=\sum_{i\alpha}C^n_{i\alpha}\ket{\phi_i\psi_\alpha}$ where $\ket{\phi_i}$ and $\ket{\psi_\alpha}$ are the orthonormal basis for the single mode and the remaining $N-1$ modes, respectively. $C^n_{i\alpha}$ is the coefficient satisfying $\sum_{i,\alpha}|C^n_{i\alpha}|^2=1$. The reduced density state of the single mode can be obtained by tracing out the rest $N-1$ modes
\begin{align}
    \rho^\prime_1=&\text{tr}_{N-1}[\rho^\prime_A]\\
    =&\sum_\gamma\braket{\psi_\gamma|\sum_nP_n\sum_{i\alpha,j\beta}C^n_{i\alpha}C^{n*}_{j\beta}\ket{\phi_i\psi_\alpha}\bra{{\phi_j\psi_\beta}}\psi_\gamma}\\
    =&\sum_nP_n\sum_{i\alpha,j\beta,\gamma}C^n_{i\alpha}C^{n*}_{j\beta}\delta_{\alpha\gamma}\delta_{\beta\gamma}\ket{\phi_i}\bra{{\phi_j}}\\
    =&\sum_nP_n\sum_{i,j,\gamma}C^n_{i\gamma}C^{n*}_{j\gamma}\ket{\phi_i}\bra{{\phi_j}}\\
    =&\sum_nP_n\sum_{i,j,\gamma}\braket{\phi_i\psi_\gamma|\Phi_n}\braket{\Phi_n|\phi_j\psi_\gamma}\ket{\phi_i}\bra{{\phi_j}}\\
    =&\frac{1}{d(E)}\sum_{i,j,\gamma}\delta_{ij}\delta_{\gamma,\gamma}\ket{\phi_i}\bra{\phi_j}=\frac{d_{N-1}(E-E_i)}{d(E)}\sum_i\ket{\phi_i}\bra{\phi_i}\label{d8},
\end{align}
where $E$ is the total system eigen-energy and $E_i$ is the eigen-energy of the single mode corresponding to $\ket{\phi_i}$. The above state Eq.~\ref{d8} is indeed thermal. To show that, note $d(E)$ and $d(E-E_i)$ are basically the number of quantum states for the total system and the $N-1$ modes, respectively. We have
\begin{equation}
    d(E)=e^{S(E)/k_B}, d_{N-1}(E-E_i)=e^{S(E-E_i)/k_B}.
\end{equation}
where the symbol $S$ denotes the entropy. Plug them into Eq.~\ref{d8}, we get
\begin{align}
    \rho_1^\prime=\sum_ie^{\frac{S(E-E_i)-S(E)}{k_B}}\ket{\phi_i}\bra{\phi_i}.
\end{align}
If we do the expansion
\begin{equation}
    S(E-E_i)=S(E)-\left(\frac{\partial S}{\partial E}\right)_{E_i=0}E_i+...,
\end{equation}
we obtain
\begin{equation}\begin{split}
    \rho^\prime_1\simeq&\sum_ie^{-\frac{1}{k_B}\left(\frac{\partial S}{\partial E}\right)_{E_i=0}E_i}\ket{\phi_i}\bra{\phi_i}\\
    =&\sum_ie^{-\beta E_i}\ket{\phi_i}\bra{\phi_i},
\end{split}
\end{equation}
which is indeed a thermal state. Following the main text, we know each mode has energy (photon number) $K\sinh^2(r)/N$. If this thermal state purified into a two-mode squeezed state, it will have a squeezing factor $r_0$ satisfying $\sinh^2(r_0)=K\sinh^2(r)/N$. The entanglement is obviously $\mathcal{E}(r_0)$, as used in the main text.


\end{appendix}

\end{document}